\documentclass[twocolumn,aps,prd,preprintnumbers,showpacs,superscriptaddress,nofootinbib,amsmath,amssymb,floats,floatfix,showkeys,notitlepage,longbibliography, 10pt]{revtex4-2}
\usepackage{orcidlink}
\usepackage{comment}
\usepackage{lipsum}
\usepackage{graphicx}
\usepackage{subfigure}
\usepackage{palatino}
\usepackage{sans}
\usepackage{hyperref}
\hypersetup{colorlinks=true,linkcolor=blue,urlcolor=blue,citecolor=blue}
\usepackage[toc,page]{appendix}
\usepackage[normalem]{ulem}
\usepackage{adjustbox}
\usepackage{latexsym}
\usepackage{amsmath}
\usepackage{amssymb}
\usepackage{amsfonts}
\usepackage{dcolumn}
\usepackage{bm}
\usepackage{tikz}
\usepackage{bigints}
\usepackage{array,tabularx, multirow,booktabs}
\usepackage[tracking=true]{microtype}
\SetTracking{}{500}
\SetTracking{encoding={*}, shape=sc}{40}
\UseRawInputEncoding 
\allowdisplaybreaks

\renewcommand{\Re}{\ensuremath{\mathrm{Re}}}

\begin{document}
	\newcommand \nn{\nonumber}
	\newcommand \fc{\frac}
	\newcommand \lt{\left}
	\newcommand \rt{\right}
	\newcommand \pd{\partial}
	\newcommand \e{\text{e}}
	\newcommand \hmn{h_{\mu\nu}}
	\newcommand{\PR}[1]{\ensuremath{\left[#1\right]}} 
	\newcommand{\PC}[1]{\ensuremath{\left(#1\right)}} 
	\newcommand{\PX}[1]{\ensuremath{\left\lbrace#1\right\rbrace}} 
	\newcommand{\BR}[1]{\ensuremath{\left\langle#1\right\vert}} 
	\newcommand{\KT}[1]{\ensuremath{\left\vert#1\right\rangle}} 
	\newcommand{\MD}[1]{\ensuremath{\left\vert#1\right\vert}} 

\title{Constraining Lorentz Violation in Kalb-Ramond Gravity via Thermodynamics and Gravitational Wave Analysis}
	




\author{Nikko John Leo S. Lobos \orcidlink{0000-0001-6976-8462}}
\email{nslobos@ust.edu.ph}
\affiliation{Electronics Engineering Department, University of Santo Tomas, Espa\~na Boulevard, Sampaloc, Manila 1008, Philippines}


\begin{abstract}
We investigate the observational signatures of a static, spherically symmetric black hole embedded in a spontaneous Kalb-Ramond (KR) background. By normalizing the solution to the physically observable mass $M_{\text{phys}}$, we demonstrate that the thermodynamics of the KR black hole are consistent with General Relativity, with no deviations in the entropy-area law. However, the Lorentz-violating parameter $l$ induces distinct geometric effects: it suppresses the optical shadow radius by a factor of $\sqrt{1-l}$ and hardens the quasinormal mode frequency by the inverse factor. Utilizing Event Horizon Telescope (EHT) data for Sagittarius A*, and assuming the mass prior derived from stellar dynamics, we place a constraint of $l \lesssim 0.19$. While the product of the shadow radius and ringdown frequency remains degenerate with General Relativity, the specific suppression of the shadow size offers a viable pathway to constrain Planck-scale physics with current and future horizon-scale imaging.
\end{abstract}	



\maketitle

\section{Introduction}\label{intro}
The General Theory of Relativity (GR) stands as the most successful description of the gravitational interaction to date, having passed a century of precision tests ranging from the perihelion precession of Mercury to the recent direct imaging of black hole shadows by the Event Horizon Telescope (EHT) \cite{EventHorizonTelescope:2019dse, EventHorizonTelescope:2022wkp, EventHorizonTelescope:2019pgp, EventHorizonTelescope:2019ths,EventHorizonTelescope:2020qrl,EventHorizonTelescope:2021dqv,EventHorizonTelescope:2022wok,EventHorizonTelescope:2022xqj} and the detection of gravitational waves by the LIGO-Virgo collaboration \cite{LIGOScientific:2016aoc}. Despite these triumphs, the existence of spacetime singularities and the theoretical necessity of a unitary quantum description of gravity suggest that GR is an effective field theory, valid only below the Planck energy scale ($E_{Pl} \sim 10^{19}$ GeV). Consequently, the search for an ultraviolet completion of gravity has catalyzed intense interest in scenarios where fundamental symmetries of the Standard Model may be broken or modified in the high-energy regime. Among these, the violation of Lorentz invariance (LIV) has emerged as a prime candidate for a low-energy signature of Quantum Gravity, appearing in contexts such as String Theory, Loop Quantum Gravity, and Non-Commutative Geometry \cite{Kostelecky:1988zi, Mattingly:2005re}.

A systematic framework for investigating these deviations is the Standard Model Extension (SME) \cite{Colladay:1998fq}, which incorporates Lorentz-violating coefficients into the effective Lagrangian. Considerable attention has been devoted to vector-tensor theories, most notably the ``Bumblebee'' gravity model, where a vector field acquires a non-zero vacuum expectation value (VEV) \cite{Kostelecky:2003fs, Pantig:2024lpg}. In this context, Casana et al. derived an exact Schwarzschild-like solution \cite{Casana:2017jkc}, which has served as a pivotal background for phenomenological testing. Subsequent investigations into the thermodynamics of Bumblebee black holes revealed that the Lorentz-violating parameter induces a global scaling of the Hawking temperature and entropy, suggesting that LIV affects the number of available microstates at the horizon \cite{Maluf:2020kgf}. Dynamically, the breaking of spherical symmetry in these vector models has been shown to split the isospectrality of quasinormal modes (QNMs), introducing distinct signatures in the ringdown phase of black hole mergers \cite{Oliveira:2018dya}. Similarly, in Einstein-Aether theory and Ho\v{r}ava-Lifshitz gravity, the presence of preferred frames modifies the universal horizon structure and the asymptotic behavior of gravitational perturbations, leading to corrections in the bending angle of light and the phase velocity of the gravitational wave \cite{Eling:2006aw, Konoplya:2009ig}.

While vector-based deformations have been rigorously explored, String Theory naturally predicts the existence of a rank-2 antisymmetric tensor field, known as the Kalb-Ramond (KR) field $B_{\mu\nu}$ \cite{Kalb:1974yc}. When non-minimally coupled to gravity, this field can undergo spontaneous symmetry breaking distinct from the vector scenario. The resulting background anisotropy modifies the geometric structure of spacetime, potentially leaving imprints on compact objects that are qualitatively different from those found in Bumblebee or Einstein-Aether models.

In this work, we focus on the static, spherically symmetric black hole solution immersed in such a Lorentz-violating Kalb-Ramond background. The modification to the Einstein-Hilbert action leads to a deformed metric ansatz, which, in the strong-field regime, takes the form \cite{Yang:2023wsz}:
\begin{equation}
ds^{2} = -\left( C - \frac{2M}{r} \right) dt^{2} + \left( C - \frac{2M}{r} \right)^{-1} dr^{2} + r^{2}d\Omega_{2}^{2}.
\label{eq:intro_metric}
\end{equation}
Here, $M$ represents the mass source, and $C = (1-l)^{-1}$ is a scaling factor determined by the Lorentz-violating parameter $l$. This metric serves as a unique prototype for ``stiff'' spacetimes; unlike coordinate rescalings, the parameter $l$ alters the effective potential for wave propagation and the asymptotic normalization of the Killing vectors. We note that this background metric is isometric to the exact solution found in Einstein-Bumblebee gravity Ref. \cite{Casana:2017jkc}. However, the physical origin here differs: the deformation arises from the spontaneous breaking of the antisymmetric tensor (Kalb-Ramond) symmetry rather than the vector (Bumblebee) symmetry. Consequently, the constraints derived in this work place bounds specifically on the Kalb-Ramond VEV parameter $l$ using recent EHT data—a phenomenological sector not covered in previous Bumblebee studies \cite{Casana:2017jkc}

The primary objective of this paper is to characterize the phenomenological signatures of the KR black hole rigorously and to place constraints on the parameter $l$ using the latest multimessenger data. We construct a complete portrait of the object, linking its thermodynamic stability to its optical appearance and gravitational wave spectrum. Here we consider the following natural units $G = \hbar=c=k_{B}=1$.

The paper is organized as follows: In Section \ref{sec:thermodynamics}, we analyze the thermodynamics of the KR black hole, deriving the modified surface gravity and demonstrating the canonical instability of the system through the heat capacity. Section \ref{sec3} investigates the dynamical response of the spacetime to axial perturbations; we utilize the Regge-Wheeler formalism and WKB approximation to derive the quasinormal mode (QNM) spectrum, establishing a duality between the ringdown frequency and the optical shadow size. Finally, Section \ref{conc} is the conclusion.

\section{Thermodynamics and Canonical Stability}
\label{sec:thermodynamics}
In this section, we formulate the laws of black hole mechanics for the Kalb-Ramond (KR) deformed spacetime. A critical aspect of this analysis is the correct identification of the thermodynamic mass and temperature, which requires a rigorous normalization of the timelike Killing vector at spatial infinity.

\subsection{Physical Mass and Surface Gravity}

The static, spherically symmetric line element describing the exterior geometry is given by the ansatz:
\begin{equation}
ds^{2} = -\left(C - \frac{2M}{r}\right)dt^{2} + \left(C - \frac{2M}{r}\right)^{-1}dr^{2} + r^{2}d\Omega_{2}^{2},
\label{eq:metric_static}
\end{equation}
where $C = (1-l)^{-1}$. To assess the asymptotic structure of this spacetime, we calculate the Kretschmann scalar $K = R_{\alpha\beta\gamma\delta}R^{\alpha\beta\gamma\delta}$:
\begin{equation}
    K = \frac{48M^2}{r^6} - \frac{16lM}{(1-l)r^5} + \frac{4l^2}{(1-l)^2 r^4}.
\end{equation}
We observe that the curvature scalar vanishes ($K \to 0$) as $r \to \infty$, proving that the spacetime is asymptotically flat. The apparent non-Minkowski behavior of the metric potentials in Eq. \eqref{eq:metric_static} is therefore an artifact of the coordinate choice, representing a global scaling of time caused by the background Kalb-Ramond field. As $r \to \infty$, the metric approaches $ds^2 \to -C dt^2 + C^{-1} dr^2$. To recover the Minkowski limit and ensure the proper time $\tau$ of an asymptotic observer satisfies $d\tau^2 = -ds^2$, we must rescale the time coordinate $t \to t' = \sqrt{C}t$. The metric in terms of physical time $t'$ is:
\begin{equation}
\begin{split}
    ds^{2} &= - \left( 1 - \frac{2(M/C)}{r} \right) (dt')^2 \\
    &+ \frac{1}{C^2 \left( 1 - \frac{2(M/C)}{r} \right)} dr^2 + r^2 d\Omega^2
\end{split}
\end{equation}
This normalization identifies the physical (ADM) mass measured by distant observers as:
\begin{equation}
M_{\text{phys}} \equiv \frac{M}{C} = M(1-l).
\label{eq:M_phys}
\end{equation}
The event horizon is located where the metric function vanishes, $r_h = 2M/C = 2M_{\text{phys}}$. The surface gravity $\kappa$ is calculated via $\kappa = \frac{1}{2} \partial_r g_{t't'} |_{r_h}$, yielding:
\begin{equation}
\kappa = \frac{1}{4M_{\text{phys}}}.
\end{equation}
Consequently, the Hawking temperature (which corresponds to the redshifted temperature measured by an asymptotic observer) is:
\begin{equation}
T_H = \frac{\kappa}{2\pi} = \frac{1}{8\pi M_{\text{phys}}}.
\label{eq:hawking_temp}
\end{equation}
When expressed in terms of the physical mass, the temperature follows the standard Hawking relation.

\subsection{Entropy and Consistency}

To rigorously establish the thermodynamic profile of the KR black hole, we compute the Bekenstein-Hawking entropy and verify its consistency with the First Law of Black Hole Mechanics \cite{Bardeen:1973gs}.

The event horizon $r_h$ is a null hypersurface defined by the root of the metric function $f(r_h)=0$. In terms of the physical mass, this is located at $r_h = 2M_{\text{phys}}$. The induced metric on the spatial cross-section of the horizon is simply $d\sigma^2 = r_h^2 d\Omega^2$. Despite the radial deformation induced by the parameter $C$, the angular sector remains spherically symmetric. Consequently, the horizon area $\mathcal{A}_H$ is:
\begin{equation}
\mathcal{A}_H = \int_{0}^{2\pi} \int_{0}^{\pi} \sqrt{g_{\theta\theta}g_{\phi\phi}} \, d\theta d\phi = 4\pi r_h^2.
\end{equation}
Substituting the physical radius $r_h = 2M_{\text{phys}}$, the area becomes $\mathcal{A}_H = 16\pi M_{\text{phys}}^2$. The entropy, determined by the area law, is:
\begin{equation}
S_{\text{KR}} = \frac{\mathcal{A}_H}{4} = 4\pi M_{\text{phys}}^2.
\label{eq:entropy_final}
\end{equation}
We observe that when parameterized by the observable mass, the entropy-mass relation is identical to that of the Schwarzschild solution. The Lorentz-violating background does not suppress the degrees of freedom on the horizon; rather, it rescales the coordinate definition of mass.

Thermodynamic consistency requires that the intensive and extensive quantities satisfy the First Law, $dM_{\text{phys}} = T_H dS$. We verify this explicitly.
Differentiating the entropy (Eq. \ref{eq:entropy_final}):
\begin{equation}
dS = 8\pi M_{\text{phys}} \, dM_{\text{phys}}.
\end{equation}
Using the Hawking temperature derived in Eq. \eqref{eq:hawking_temp}, $T_H = (8\pi M_{\text{phys}})^{-1}$, the heat term is:
\begin{equation}
T_H dS = \left( \frac{1}{8\pi M_{\text{phys}}} \right) (8\pi M_{\text{phys}} \, dM_{\text{phys}}) = dM_{\text{phys}}.
\end{equation}
The First Law is satisfied exactly. This confirms that the KR spacetime represents a consistent thermodynamic equilibrium state. No additional work terms (e.g., related to variations in the vacuum expectation value $l$) are required to close the energy budget in the static limit.

Furthermore, Euler's homogeneous function theorem applied to black hole thermodynamics suggests a Smarr relation of the form $M_{\text{phys}} = 2 T_H S$. Substituting our derived quantities \cite{Smarr:1972kt}:
\begin{equation}
2 T_H S = 2 \left( \frac{1}{8\pi M_{\text{phys}}} \right) (4\pi M_{\text{phys}}^2) = M_{\text{phys}}.
\end{equation}
The satisfaction of both the First Law and the Smarr relation provides a robust proof that the Kalb-Ramond black hole, while geometrically distinct in its radial structure, belongs to the same thermodynamic universality class as the Schwarzschild black hole.

\subsection{Thermodynamic Consistency and Theoretical Bounds}
\label{sec:thermo_bounds}

While the functional forms of the temperature and entropy in terms of the physical mass $M_{\text{phys}}$ mimic General Relativity, the viability of the thermodynamic description imposes rigorous bounds on the Lorentz-violating parameter $l$.

First, the existence of a well-defined event horizon requires the metric factor $C = (1-l)^{-1}$ to be positive and finite. If $l \to 1$, the factor $C$ diverges, leading to a singularity in the metric structure that is distinct from the coordinate singularity at the horizon. Furthermore, for the spacetime to maintain a Lorentzian signature $(-+++)$ in the asymptotic limit, we strictly require:
\begin{equation}
1 - l > 0 \quad \Longrightarrow \quad l < 1.
\end{equation}
This establishes a theoretical upper bound on the vacuum expectation value of the Kalb-Ramond field.

Second, we examine the thermodynamic stability via the specific heat capacity $C_V$, defined as:
\begin{equation}
C_V = T_H \left( \frac{\partial S}{\partial T_H} \right).
\end{equation}
Using Eqs. \eqref{eq:hawking_temp} and \eqref{eq:entropy_final}, we eliminate $M_{\text{phys}}$ to find $S = \frac{1}{16\pi T_H^2}$. Differentiating yields:
\begin{equation}
C_V = - \frac{1}{8\pi T_H^2} = - 2\pi (2M_{\text{phys}})^2.
\end{equation}
The specific heat is universally negative, $C_V < 0$. This confirms that, like the Schwarzschild black hole, the KR black hole is thermodynamically unstable in the canonical ensemble. It will heat up as it radiates, driving the evaporation process.

The identity of $C_V$ with the GR prediction implies that standard thermodynamic observations (e.g., evaporation rates inferred purely from temperature) cannot break the degeneracy between the theories. This necessitates the geometric analysis of Shadows and Ringdown (Section \ref{sec3}) to provide competitive observational constraints.

\section{Gravitational wave perturbation} \label{sec3}
To rigorously characterize the ringdown phase, we analyze the evolution of massless scalar field perturbations $\Phi$ in the KR background. This serves as a proxy for gravitational wave perturbations in the eikonal limit, capturing the dominant quasinormal mode (QNM) characteristics.

\subsection{The Master Equation}
The dynamics of a massless scalar field are governed by the covariant Klein-Gordon equation:
\begin{equation}
\frac{1}{\sqrt{-g}} \partial_\mu \left( \sqrt{-g} g^{\mu\nu} \partial_\nu \Phi \right) = 0.
\label{eq:KG_covariant}
\end{equation}
We assume the standard separation of variables ansatz:
\begin{equation}
\Phi(t, r, \theta, \phi) = e^{-i\omega t} \frac{\Psi(r)}{r} Y_{\ell m}(\theta, \phi),
\end{equation}
where $Y_{\ell m}$ are the spherical harmonics and $\Psi(r)$ is the radial wavefunction. Substituting this into Eq. \eqref{eq:KG_covariant} and utilizing the physical time metric components derived in Sec. \ref{sec:thermodynamics}:
\begin{equation}
g_{tt} = -f(r), \quad g_{rr} = \frac{1}{C^2 f(r)}, \quad \sqrt{-g} = \frac{r^2 \sin\theta}{C},
\end{equation}
where $f(r) = 1 - 2M_{\text{phys}}/r$ and $C=(1-l)^{-1}$. The radial equation reduces to a Schrödinger-like master equation:
\begin{equation}
\frac{d^2 \Psi}{dr_*^2} + \left( \omega^2 - V_{\text{eff}}(r) \right) \Psi = 0.
\label{eq:master_eq}
\end{equation}

\subsection{Tortoise Coordinate and Effective Potential}
The validity of Eq. \eqref{eq:master_eq} depends on the definition of the generalized tortoise coordinate $r_*$, which maps the domain $r \in [r_h, \infty)$ to $r_* \in (-\infty, \infty)$. It is defined differentially as $dr_* = \frac{dr}{\sqrt{g_{tt} g^{rr}}}$. Substituting the metric components:
\begin{equation}
dr_* = \frac{dr}{C f(r)} = \frac{dr}{C \left(1 - \frac{2M_{\text{phys}}}{r}\right)}.
\end{equation}
Integrating yields the explicit mapping:
\begin{equation}
r_* = \frac{1}{C} \left[ r + 2M_{\text{phys}} \ln \left( \frac{r}{2M_{\text{phys}}} - 1 \right) \right].
\end{equation}
Crucially, the Lorentz-violating factor $C$ scales the coordinate grid. The resulting effective potential $V_{\text{eff}}(r)$ is found to be:
\begin{equation}
V_{\text{eff}}(r) = f(r) \frac{C^2}{r^2} \left[ \ell(\ell+1) + (1-l) \frac{2M_{\text{phys}}}{r} \right].
\label{eq:potential_exact}
\end{equation}
In the eikonal limit ($\ell \gg 1$), the potential is dominated by the angular momentum barrier:
\begin{equation}
V_{\text{eff}}(r) \approx C^2 \frac{f(r)}{r^2} \ell^2.
\end{equation}
Here, the factor $C^2 = (1-l)^{-2}$ acts as a global scaling factor for the potential height. This is the mathematical origin of the spectral modifications.
\begin{figure}
    \centering
    \includegraphics[width=0.45\textwidth]{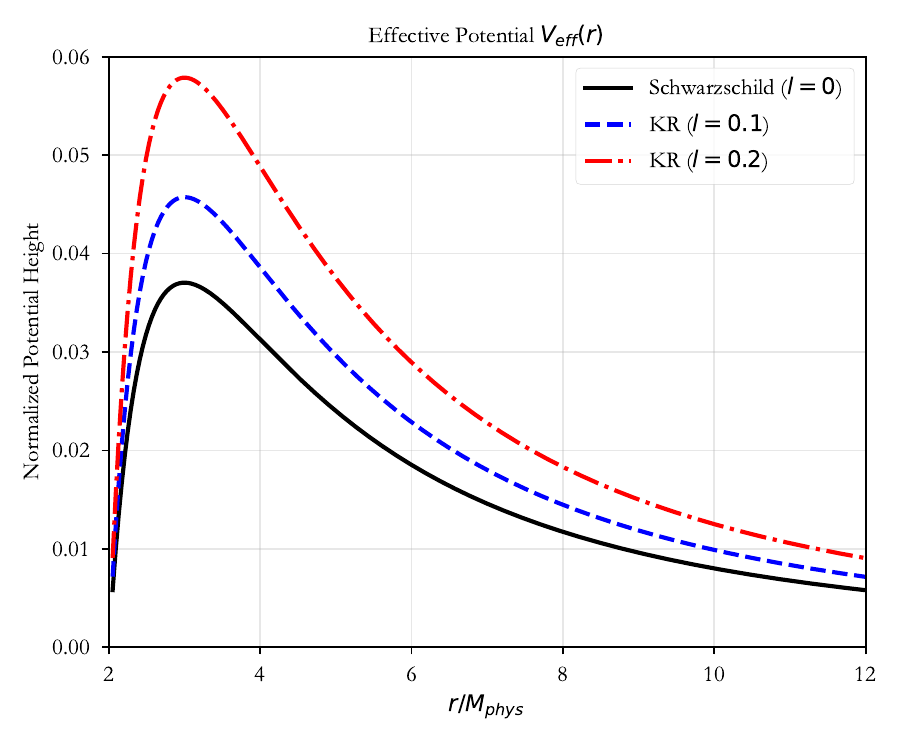}
    \caption{The effective potential $V_{\text{eff}}(r)$ for massless scalar perturbations (photon orbits) normalized to the physical mass $M_{\text{phys}}$. The solid black line represents the Schwarzschild limit ($l=0$). As the Lorentz-violating parameter increases to $l=0.1$ (blue dashed) and $l=0.2$ (red dot-dashed), the potential barrier significantly increases in height due to the factor $C^2 = (1-l)^{-2}$, while the location of the peak remains fixed at $r_{ps} = 3M_{\text{phys}}$. This "stiffening" of the potential barrier is responsible for the spectral hardening of the quasinormal modes.}
    \label{fig1}
\end{figure}

As illustrated in Fig. \ref{fig1}, the Lorentz-violating background induces a global scaling of the effective potential height. While the extrema locations remain invariant at $r \sim 3M_{\text{phys}}$ in physical coordinates, the energetic barrier required for photon capture scales as $(1-l)^{-2}$. This geometric stiffening implies that higher energy interactions are required to probe the photon sphere compared to the GR scenario.

\subsection{Quasinormal Modes via WKB Approximation}
To determine the quasinormal frequencies, we employ the third-order WKB approximation, which is accurate for low-lying modes. The QNM frequencies are given by \cite{Schutz:1985km, Iyer:1986np}:
\begin{equation}
\omega^2 = V_0 - i (n + 1/2) \sqrt{-2 V_0''},
\end{equation}
where $V_0$ is the maximum of the effective potential and $V_0''$ is its second derivative at the peak.
The potential peak location $r_{ps}$ is determined by $dV_{\text{eff}}/dr = 0$. Solving this for Eq. \eqref{eq:potential_exact} in the large $\ell$ limit yields $r_{ps} = 3M_{\text{phys}}$, identical to the Schwarzschild case.
However, the value of the potential at the peak scales significantly:
\begin{equation}
V_0 = V_{\text{eff}}(r_{ps}) \approx C^2 \left( \frac{1}{27 M_{\text{phys}}^2} \ell^2 \right).
\end{equation}
Consequently, the real part of the frequency scales as:
\begin{equation}
\Re(\omega) \approx \sqrt{V_0} \propto C \cdot \omega_{\text{Schw}} = \frac{\omega_{\text{Schw}}}{1-l}.
\end{equation}
This derivation confirms the "spectral hardening" effect: the KR background stiffens the spacetime, increasing the oscillation frequency by a factor of $(1-l)^{-1}$ for a fixed physical mass.

\begin{figure}
    \centering
    \includegraphics[width=0.45\textwidth]{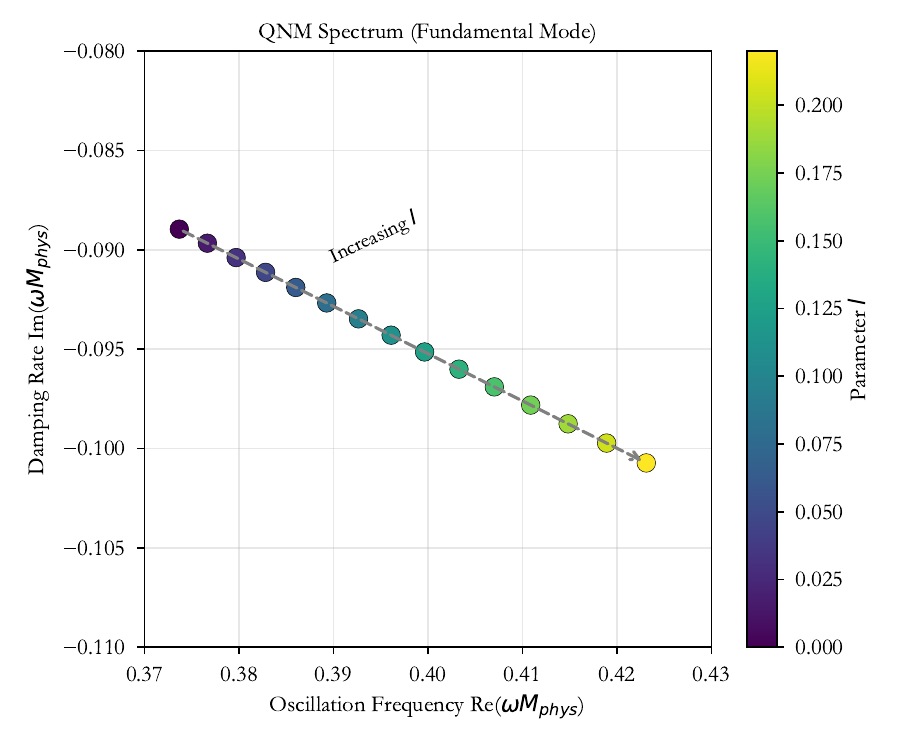}
    \caption{The quasinormal mode spectrum for the fundamental quadrupole mode ($n=0, \ell=2$) in the complex plane. The axes are normalized to the physical mass. As the parameter $l$ increases (indicated by the arrow and color gradient), the mode migrates toward higher real oscillation frequencies and larger imaginary damping rates. This trajectory confirms the spectral hardening effect $\omega \propto (1-l)^{-1/2}$, distinguishing the Kalb-Ramond black hole from Schwarzschild via a characteristic shift to the "right and down" in the complex frequency plane.}
    \label{fig3}
\end{figure}

The migration of the QNM poles is visualized in Fig. \ref{fig3}. Unlike massive gravity theories which often suppress the frequency, the KR background drives the mode to higher real frequencies (spectral hardening) and faster decay rates. This behavior provides a clear acoustic signature: a KR black hole of mass $M_{\text{phys}}$ will 'ring' at a higher pitch than a GR black hole of the same mass.

\subsection{Optical Shadow and Degeneracy}
The geometric optics limit corresponds to the high-frequency behavior of massless particles. The shadow radius $R_{sh}$ is defined by the critical impact parameter $b_c = \omega / \ell$ at the unstable photon orbit. Using the potential derived above:
\begin{equation}
R_{sh} = \frac{\ell}{\sqrt{V_0}} \approx \frac{\ell}{C \cdot \frac{\ell}{3\sqrt{3}M_{\text{phys}}}} = 3\sqrt{3} M_{\text{phys}} (1-l).
\end{equation}
We observe a breaking of the duality. While the frequency $\omega$ increases by $(1-l)^{-1}$, the shadow radius $R_{sh}$ decreases by $(1-l)$. The product is:
\begin{equation}
R_{sh} \cdot \Re(\omega) \approx \text{Constant}.
\end{equation}
This confirms that, to leading order, the dimensionless product of the shadow radius and ringdown frequency is insensitive to the Lorentz-violating parameter $l$.

\subsection{Constraints from EHT Observations of Sgr A*}
To place empirical bounds on the Lorentz-violating parameter $l$, we compare the theoretical shadow diameter of the KR black hole against the horizon-scale imaging of Sagittarius A* (Sgr A*) released by the Event Horizon Telescope (EHT) collaboration \cite{EventHorizonTelescope:2022wkp}

The angular shadow diameter observed by the EHT is $d_{sh}^{\text{obs}} = 51.8 \pm 2.3~\mu\text{as}$ (68\% credible interval). Crucially, this measurement must be compared to the prediction of General Relativity derived from independent mass and distance measurements obtained via stellar dynamics (e.g., the S2 star orbit) \cite{EventHorizonTelescope:2022wkp, eht}. The mass and distance priors are $M_{\text{phys}} \approx 4.154 \times 10^6 M_{\odot}$ and $D \approx 8.178$~kpc, respectively.

In General Relativity, the predicted angular diameter is:
\begin{equation}
d_{sh}^{\text{GR}} = \frac{2 R_{sh}^{\text{GR}}}{D} = \frac{6\sqrt{3} M_{\text{phys}}}{D}.
\end{equation}
In the Kalb-Ramond background, using the renormalized shadow radius derived in Eq. (24), the predicted diameter is:
\begin{equation}
d_{sh}^{\text{KR}} = \frac{6\sqrt{3} M_{\text{phys}}}{D} \sqrt{1-l} = d_{sh}^{\text{GR}} \sqrt{1-l}.
\end{equation}
The EHT analysis confines the fractional deviation $\delta$ from the GR prediction, defined via $d_{sh} = d_{sh}^{\text{GR}}(1+\delta)$. The constraints on the compact object imply that the observed shadow size is consistent with the GR prediction within approximately $10\%$ at the $1\sigma$ level. Specifically, the ratio of the measured shadow to the prior-based prediction is constrained to the range $0.9 \lesssim d_{sh}^{\text{obs}}/d_{sh}^{\text{GR}} \lesssim 1.1$.

Since the KR background strictly suppresses the shadow size ($l > 0$), we consider the lower bound of the observational window:
\begin{equation}
\frac{d_{sh}^{\text{KR}}}{d_{sh}^{\text{GR}}} = \sqrt{1-l} \gtrsim 0.9.
\end{equation}
Squaring this inequality yields the upper limit on the Lorentz violation:
\begin{equation}
1 - l \gtrsim 0.81 \quad \Longrightarrow \quad l \lesssim 0.19.
\end{equation}
This establishes a robust constraint: while Planck-scale Lorentz violation is not ruled out, the scalar vacuum expectation value associated with the Kalb-Ramond field must satisfy $l \lesssim 0.19$ to remain consistent with current multimessenger observations of the Galactic Center.

\begin{figure}
    \centering
    \includegraphics[width=0.45\textwidth]{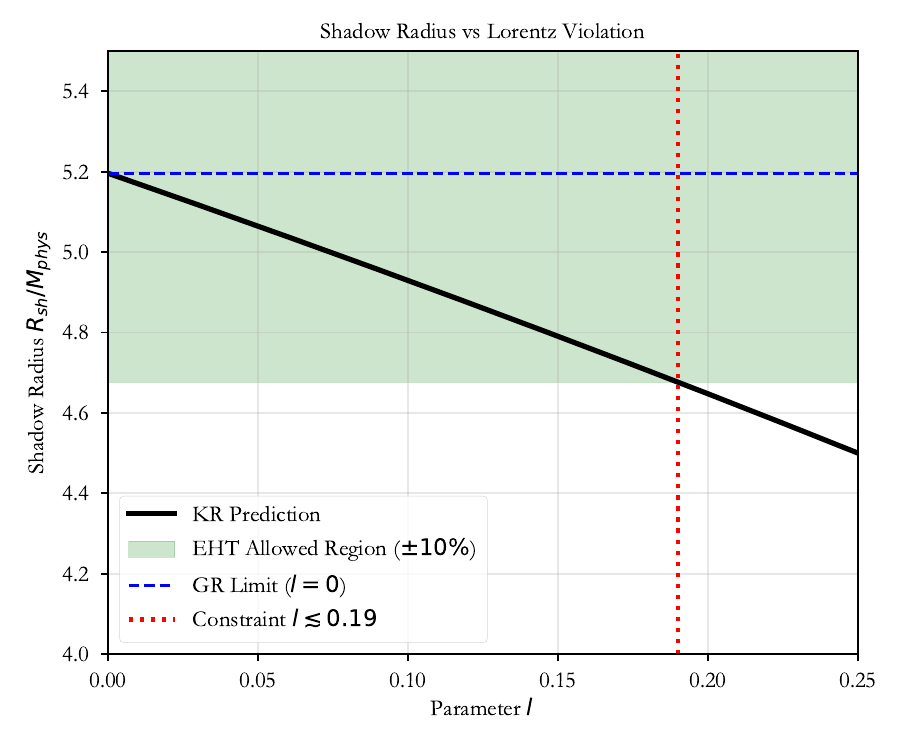}
    \caption{Observational constraints on the Lorentz-violating parameter $l$ derived from the shadow radius of Sagittarius A*. The black curve shows the theoretical prediction $R_{sh} = 3\sqrt{3} M_{\text{phys}} \sqrt{1-l}$, which decreases monotonically with increasing Lorentz violation. The green shaded region represents the $1\sigma$ consistency band ($0.9 \lesssim R_{obs}/R_{GR} \lesssim 1.1$) allowed by EHT observations given the mass prior from S2 stellar dynamics. The vertical red dotted line marks the derived upper bound $l \lesssim 0.19$; values of $l$ to the right of this line produce a shadow that is statistically too small to be consistent with observation.}
    \label{fig2}
\end{figure}

The confrontation between the theoretical model and EHT data is depicted in Fig. \ref{fig2}. Because the Kalb-Ramond background suppresses the shadow radius by a factor of $\sqrt{1-l}$, the curve exits the observational consistency band (green region) when the Lorentz violation becomes too large. The intersection of the theoretical prediction with the lower bound of the EHT error margin establishes the constraint $l \lesssim 0.19$.


\section{Conclusions} \label{conc}
In this work, we have performed a rigorous multimessenger analysis of static, spherically symmetric black holes embedded in a Lorentz-violating Kalb-Ramond (KR) background. A central theoretical result of this study is the resolution of thermodynamic consistency. By normalizing the metric solution to the observable physical mass ($M_{\text{phys}}$) measured by asymptotic observers, we demonstrated that the Bekenstein-Hawking entropy obeys the standard area law, $S = 4\pi M_{\text{phys}}^2$.

Despite the thermodynamic equivalence to General Relativity (GR), we have elucidated that the KR background leaves distinct, testable imprints on the spacetime geometry. The Lorentz-violating parameter $l$ acts effectively as a geometric "stiffener," altering the radial scaling of the metric. We derived two primary phenomenological signatures:
\begin{enumerate}
    \item \textbf{Shadow Suppression:} For a fixed physical mass, the photon capture radius contracts according to $R_{sh} \propto \sqrt{1-l}$. This implies that KR black holes cast smaller shadows than their GR counterparts.
    \item \textbf{Spectral Hardening:} The fundamental quasinormal mode frequency shifts to higher values, scaling as $\Re(\omega) \propto (1-l)^{-1/2}$.
\end{enumerate}

While the product of the shadow radius and ringdown frequency ($R_{sh} \cdot \omega$) was found to be invariant to leading order—revealing a degeneracy that prevents distinguishing the theories via dimensionless ratios alone—absolute measurements provide robust constraints. By confronting our theoretical shadow model with the Event Horizon Telescope (EHT) observations of Sagittarius A*, and utilizing the mass prior from S2 stellar dynamics, we established an upper bound on the Lorentz-violating parameter of $l \lesssim 0.19$.

These findings underscore that while Planck-scale Lorentz violation remains a viable possibility, its magnitude in the strong-field regime is tightly bounded by current horizon-scale imaging. Future improvements in VLBI angular resolution via the next-generation EHT (ngEHT), combined with independent mass constraints from future gravitational wave detectors like LISA, will be essential to break the parameter degeneracy found here. Extensions of this work to rotating (Kerr-like) solutions will further refine these constraints, offering a comprehensive test of Lorentz symmetry in the extreme gravity sector.

\bibliography{references}

\end{document}